\documentclass[a4paper]{jpconf}
\usepackage{graphicx}
\usepackage{xcolor}
\usepackage{amsmath}
    \newcommand{\un}[1]{\,\mathrm{#1 }}

    \renewcommand{\e}{\mathrm{e}}
     
    \def\A{\ensuremath{\text{\AA}}} 
    \newcommand{\up}{{\left|+\right>}}
    \newcommand{\dn}{{\left|-\right>}}
\usepackage{amssymb}

\begin{document}
\title{Efficient polarization analysis for focusing neutron instruments}

\author{Jochen Stahn$^1$, Artur Glavic$^1$}

\address{$^1$ Laboratory for Neutron Scattering and Imaging, Paul Scherrer Institut, Villigen PSI, Switzerland}

\ead{artur.glavic@psi.ch}

\begin{abstract}
  Polarized neutrons are a powerful probe to investigate magnetism in condensed matter on length scales from single atomic distances to micrometers.
  With the ongoing advancement of neutron optics, that allow to transport beams with increased divergence, the demands on neutron polarizes and analyzers have grown as well.
  The situation becomes especially challenging for new instruments at pulsed sources, where a large wavelength band needs to be polarized to make efficient use of the time structure of the beam.
  
  Here we present a polarization analysis concept for highly focused neutron beams that is based on transmission supermirrors that are bend in the shape of equiangular spirals.
  The method allows polarizations above 95\% and good transmission, without negative impact on other beam characteristics.
  An example of a compact polarizing device already tested on the AMOR reflectometer is presented as well as the concept for the next generation implementation of the technique that will be installed on the Estia instrument being build for the European Spallation Source.
\end{abstract}

\section{Introduction}
  Neutron radiation has several properties desired in the investigation of many condensed matter systems.
  In addition to a suitable wavelength on the \A{} length scale, high sensitivity to lighter elements and energies useful for spectroscopic studies, the main advantage the neutrons pose over x-ray photons is the sensitivity to magnetic induction inherent to the magnetic moment of the particle.
  The neutron is a spin $1/2$ particle and thus has only two allowed values for the z-component of its magnetic moment, aligned parallel or anti-parallel to an external magnetic field. 
  One speaks of spin-up/-down polarization, denoted by $\left|+\right>$/$\left|-\right>$.

  Although interaction with magnetic moments in the investigated sample alone can reveal information about the magnetic structure, e.g. magnetic unit cell, symmetry and local moment from powder diffraction, the determination of the neutron polarization before and after the scattering process yields additional knowledge about the direction of magnetization\cite{Halpern1939,Blume1963,Lovesey1986}.
  There are basically two principles for the interaction of polarized neutrons with magnetic moments in the sample.
  Only the projection $\vec{M}_{\perp}$ of the magnetization vector $\vec{M}$ onto the plane perpendicular to the scattering vector $\vec{Q}$ is detected.
  Secondly, the relative orientation of this projection to the neutron polarization vector $\vec{P}$ determines the strength of the interaction and change of polarization state:
  
  \begin{eqnarray}
    \text{non spin-flip  scattering: }& \vec{M}_{\perp}|| \vec{P}\\
    \text{spin-flip  scattering: }& \vec{M}_{\perp}\perp \vec{P}
  \end{eqnarray}

  In the first case the scattering potential is increased in the parallel and decreased in the anti-parallel case.

  Neutron beams can essentially be polarized using one of three methods:
  \begin{itemize}
   \item Magnetic Bragg-scattering in special Heusler alloy crystals, where the nuclear and magnetic cross-sections balance out in the structure factor to only allow one spin state to get reflected\cite{Nathans1959,Kulda2001,Takeda2008,Groitl2015}. 
        The practical advantage of this technique is the simple implementation at instruments that already use crystal monochromators without the need for extra space or complicated arrangements.
        Polarization efficiency, however, is rather limited and the technique is not suitable for polychromatic beams.
   \item The spin dependent absorption cross-section of $^3$He can be used to polarize neutrons by generating a partially nuclear spin polarized gas volume where the neutron beam is transmitted through \cite{Heil1999,Chen2007,Parnell2015}.
        The much larger attenuation of the undesired spin-state results in a polarized neutron beam.
        Main advantages of this method are the possibility to specifically optimize the polarization efficiency to the desired application and the relatively low dependence of the device properties on the neutron trajectory.
        Especially the latter can be used for polarization analysis with large angular coverage.
        A main limitation of $^3$He polarizers is the sensitivity to magnetic field perturbation that make it difficult to use them near high-field magnets.
        Secondly, the wavelength dependence of the absorption process leads to significant losses in transmission for longer wavelength, especially relevant for instruments with large wavelength bands.
        The transmission over the bandwidth of a time of flight instrument can vary by a factor of 5 or larger.
   \item Reflection of an artificial multilayer or supermirror with different reflectivities for spin-up and spin-down neutrons\cite{Mesei1976,Stahn2002,Padiyath2006}.
        These devices can use varying mirror geometries adapted to the specific application, but the physical principle is the same.
        While limited by the small reflection angle needed for each mirror, these solid state devices can reach very high neutron polarizations with only minimal losses to the intensity of the desired neutron polarization state.
        In environments with constrained space the necessary system size and/or the high magnetic field needed to saturate the magnetic coating can lead to interference with other instrument components.
  \end{itemize}
  
  Here we present a system based on the latter method.
  The fundamental properties of reflecting polarization mirrors are:
  \begin{itemize}
   \item Reflectivity $R$ is a function of the momentum transfer normal to the reflecting surface $q_z = \frac{4\pi}{\lambda} \sin \alpha$ with the neutron wavelength $\lambda$ and the incident angle $\alpha$.
   \item The effective upper critical $q_z$ for the two spin states are
    $q_c^{\left|+\right>} := m~ q_c^\mathrm{Ni}$ and 
    $q_c^{\left|-\right>}$, where the critical $q_z$ of natural Ni 
    $q_c^\mathrm{Ni}\approx 0.0022\un{\text{\A}^{-1}}$ is used as a reference (m=1).
   \item The polarization efficiency of the polarizer for the reflected beam is
    \begin{eqnarray}
      P_R &=& \frac{R^{\left|+\right>} - R^{\left|-\right>}}{R^{\left|+\right>} + R^{\left|-\right>}} \\
          &\approx& 1-\frac{R^{\left|+\right>}}{R^{\left|-\right>}} \quad
            \mbox{ for } q_c^{\left|-\right>} < q_z < q_c^{\left|+\right>}.
    \end{eqnarray}
    Here $R^{\left|-\right>} \ll R^{\left|+\right>}$ was used.
    Typical polarizers based on Fe/Si or FeCoV/Ti:N supermirrors (SM) reach $P = 92\%\dots99\%$ with a single reflection.
   \item For the transmitted beam the situation is different: 
    The transmission $T^{\left|-\right>} = 1-R^{\left|-\right>}$ is close to 100\% for $q_z > q_c^\dn$, but $T^{\left|+\right>} = 1-R^{\left|+\right>}$ can reach up to 50\% for high $m$. 
    After a single transmission the polarizing efficiency
    \begin{eqnarray}
      P_T &=& \frac{T^{\left|-\right>} - T^{\left|+\right>}}{T^{\left|+\right>} + T^{\left|-\right>}} 
    \end{eqnarray}
    thus is in the range 30\% to 90\%.

    For a good polarization, several SM coatings must be used in sequence. 
    Often the supporting substrate (a Si waver) is coated on both sides, and sometimes a stack of substrates is used.
  \end{itemize}
  
  The advantage of using the transmitted beam is, that its trajectory is hardly affected by the polarizer. 
  In addition such devices can be used to suppress long wavelength contamination from frame overlap in pulsed applications, as the transmission for both spin states below the critical edge of the substrate is 0.
  This manuscript deals essentially with a special geometry of a SM used in transmission that allows polarization of neutron beams with large divergence.
  We make use of the small focus size in modern, highly focused neutron instruments that provide the necessary boundary condition, namely a limited local angular distribution of neutron trajectories.
  
\section{Basic principle}
  For a given angular distribution $\Delta \alpha$ and wavelength spread $[\lambda_\mathrm{min}, \lambda_\mathrm{max}]$ the necessary $m$ of the coating can be calculated with
  \begin{eqnarray}
    q_c^{\left|-\right>} &<& \frac{4\pi}{\lambda_\mathrm{max}}  \sin \alpha_\mathrm{min} \nonumber \\
    \alpha_\mathrm{min} &>& \arcsin \frac{q^{\left|-\right>} \lambda_\mathrm{max}}{4\pi}
      \label{eqn:alphais}
  \end{eqnarray}
  and
  \begin{eqnarray}
    m \, q_c^\mathrm{Ni} &\ge & \frac{4\pi}{\lambda_\mathrm{min}} \sin{(\alpha_\mathrm{min}+\Delta\alpha)}
      \label{eqn:mis}
    \quad ,
  \end{eqnarray}
  This limits the application of SM polarizers to cases where either the $\alpha$- or the $\lambda$-range are moderate or where a lot of space is available (due to the small angles of incidence $\alpha$ on the polarizer). 
  The latter is the case if a (part) of the neutron beam guide is coated with a polarizing SM or if a polarizing cavity\cite{Krist1992} is used. 

  A transmission filter used for polarizing a beam or for frame-overlap suppression (i.e. used as a low-pass for $\lambda$) should ideally intersect the beam at the same optimized angle $\hat\alpha$ everywhere within the beam. 
  For a parallel beam this is fulfilled by a flat surface, inclined relative to the beam by $\hat\alpha$.
  Although a highly beam divergence ($\Delta \theta$) seems to be incompatible with this condition on the first glance, the knowledge of a point of origin/convergence is enough to generate a curved surface that intersects with the beam under the same angle over the whole divergence range.
  The corresponding surface has the shape of a logarithmic spiral, also called \textsl{equiangular} spiral.

\section{Features of an equiangular spiral}
  The general parametric representation of an equiangular spiral is
   \begin{eqnarray*}
     x &=& a\,\e^{b\theta}\,\cos \theta \\
     y &=& a\,\e^{b\theta}\,\sin \theta 
   \end{eqnarray*}
  where $a$ is a scaling factor and $b = 1/\tan \hat\alpha$ relates to the intersection angle $\hat\alpha$ of the spiral with an arbitrary trajectory $y=x\,\tan\theta$ through the pole at $x=y=0$. 

  A feature of the equiangular spiral is that a central stretching, i.e.\ changing $a$ is equivalent to a rotation $\theta \rightarrow \theta+\theta'$ around the pole.

  For a small $\hat\alpha$ and a rather low divergence $\Delta\theta \ll 10^\circ$ the 
  spiral can be approximated by the function 
  \begin{eqnarray} 
    f(x) &=& \hat\alpha \, x \, \ln\left[\,\frac{x}{x_{min}}\,\right]
  \end{eqnarray}
  where $x_{min}$ is a scaling factor 
  given by the intersection of $f(x)$ with the horizon $f(x_{min}) = 0$.

  \begin{figure}[t]
    \centering
    \includegraphics[width=8cm]{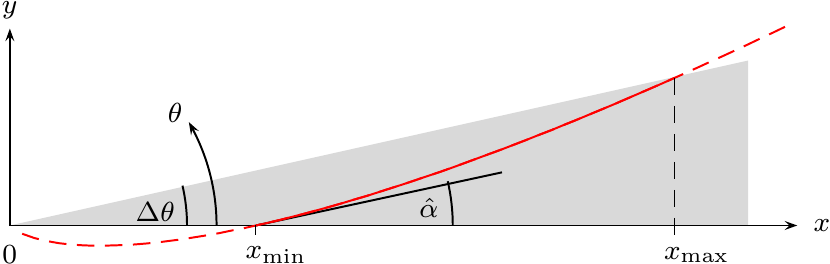}
   \caption{\label{fig:xlnx}
     Geometry of the equiangular spiral polarizer in its scattering plane. 
     The dashed line is the spiral, the solid part the actually realized transmission reflector covering the beam (gray area) emerging from the spiral pole.
   }
  \end{figure}

\section{Single lamella device}
  The simplest realization of a polarization- or wavelength filter using the presented concept is a single face with spiral bending in one direction  (illustrated in figure \ref{fig:xlnx}) and no bending in the perpendicular direction. 
  Such a device can in principle cover arbitrary divergences $\Delta\theta$, where the length of the device growth exponentially with $\Delta\theta$.

  The optimization of the device, i.e.\ the chioce of $\hat\alpha$ and $m$, has to take into account the final width of the source $\pm \Delta y$. 
  This leads to angular errors at the polarizer of $\Delta\alpha= \arcsin \Delta y / x$ largest for the point closest to the pole $x_{min}$. 
  Thus eqns. (\ref{eqn:alphais}) and (\ref{eqn:mis}) are replaced by
  \begin{eqnarray}
    \hat{\alpha} &>& \arcsin\frac{q_c^\dn \,\lambda_\mathrm{min}}{4\pi}
                     + \arcsin\frac{\Delta y}{x_{min}}
  \end{eqnarray}
  and
  \begin{eqnarray}
    m &>& \frac{4\pi}{q_c^\mathrm{Ni} \, \lambda_\mathrm{min}} 
          \left( \sin \hat{\alpha} + \frac{\Delta y}{x_{min}}\right)
  \end{eqnarray}

 \subsection{Demonstrator device}
    A polarizer based on this principle was designed and build for the Selene demonstrator at the AMOR reflectometer of PSI\cite{Stahn2016}.
    The device was designed at the Laboratory for Neutron Scattering and manufactured by SwissNeutronics.
    The rather limited space of $40\un{cm}$ from source to guide entrance and the relatively large source with a height of $2\Delta y = 1\un{mm}$ necessitate a split assembly, i.e.\ the device has a V-shape comparable to a polarizing V-cavity\cite{Krist1992}.%
    \footnote{ Despite the similar shape, this is no cavity. The polarization is based on exactly one interaction of the beam with the polarizing reflector, there are no outer (reflecting) walls.} 
  \begin{figure}[t]
    \centering
    \includegraphics[width=0.35\textwidth]{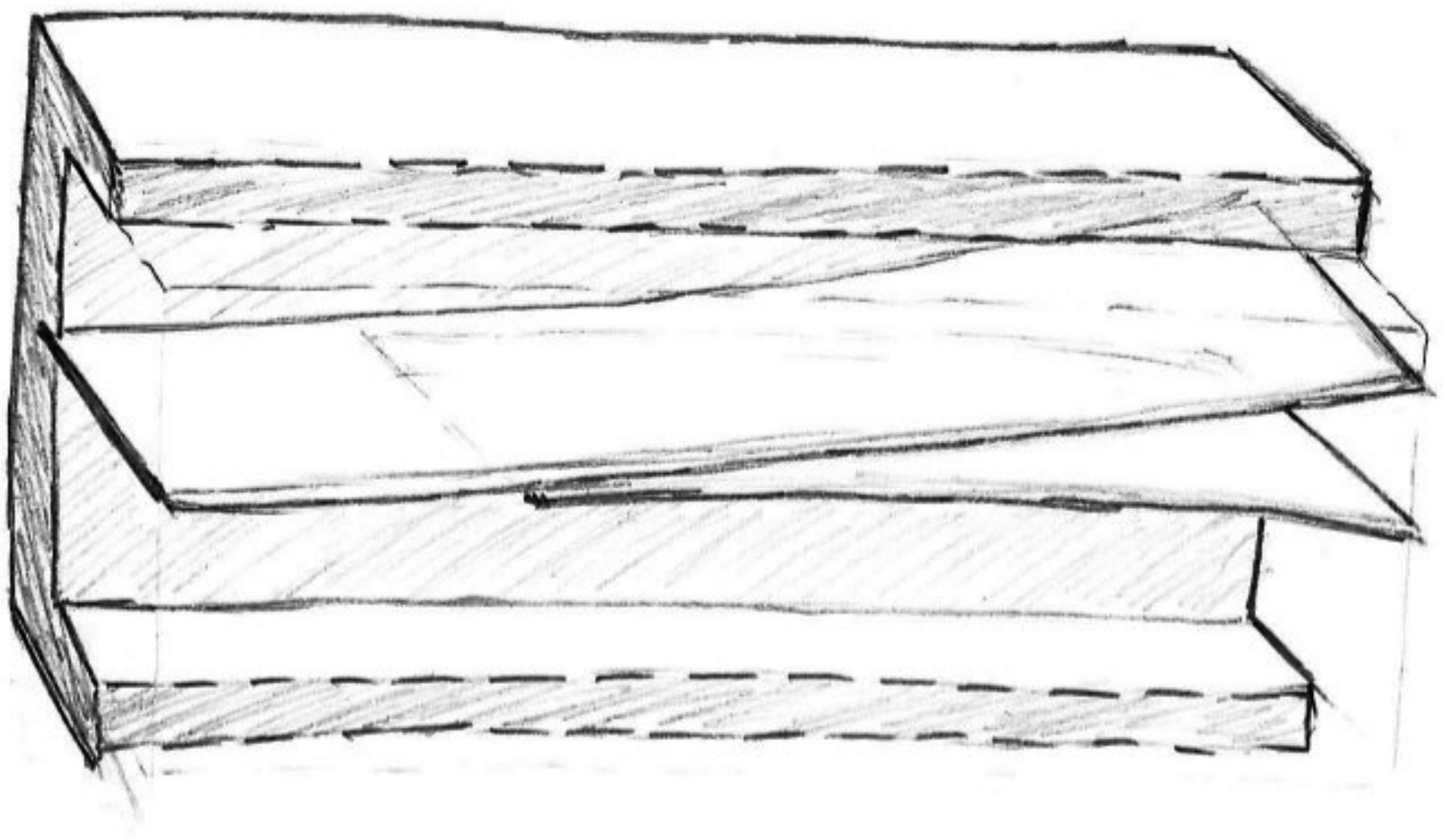} \hspace*{5mm}
    \includegraphics[width=0.6\textwidth]{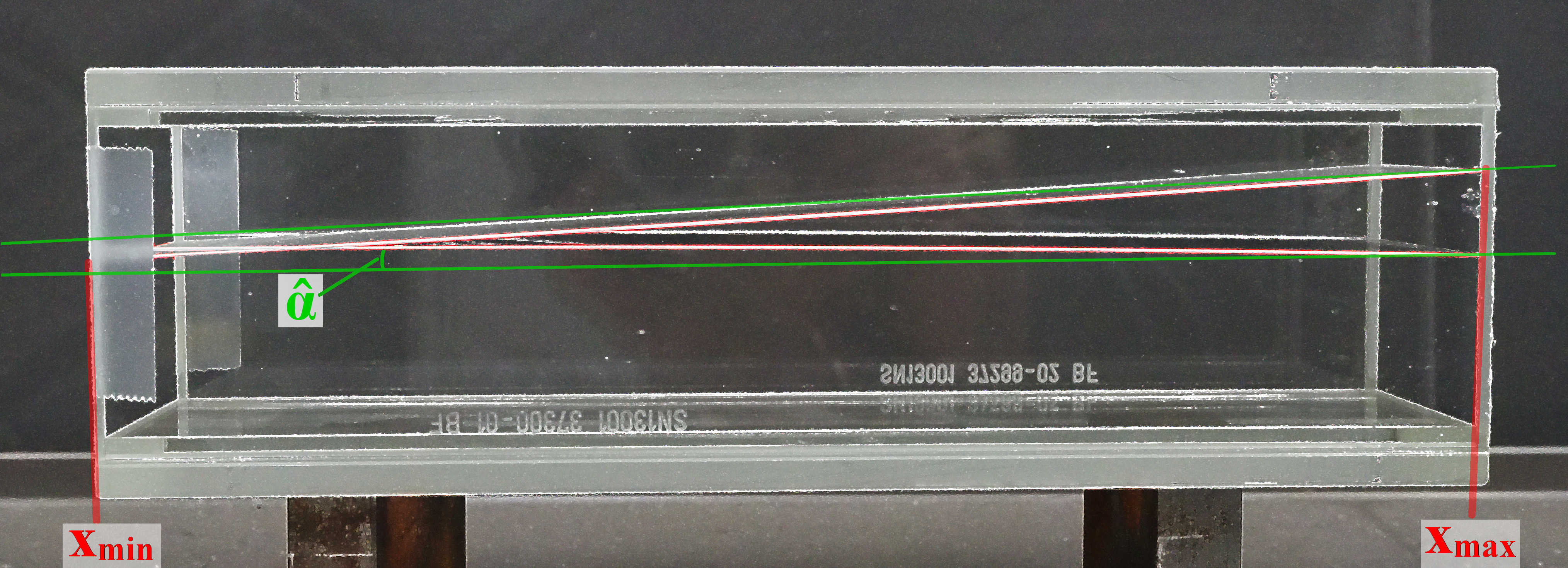}
    \caption{\label{fig:optics_xlnxphoto}
      Left: First sketch for a frame-overlap filter and polarizer for the Selene prototype set-up, based on transmission/reflection through/at spiral-curved surfaces.
      Right: Final device, produced by SwissNeutronics. 
      The glass box is $215\un{mm}$ long and has an inner height and width of $50\times50\un{mm^2}$. 
      The virtual source is to be $170\un{mm}$ before the entrance (to the left).
    }
  \end{figure}
  
  Figure \ref{fig:optics_xlnxphoto} shows the first design sketch and the final device.
  The technical parameters of the latter are $\lambda \in [4, 16]\un{\A}$, $x_{min} = 170\un{mm}$ and $x_{max} = 395\un{mm}$ (point furthest from the pole).
  The divergence to be covered is $\Delta\theta = 1.8^\circ$. 
  The angle of incidence is $\hat\alpha = 1.5^\circ$ and the polarizing coating is a $m=4.2$ FeCoV / Ti:N supermirror, covered with Ni to simultaneously act as a low-pass for $\lambda < 16.8\un{\A}$, i.e.\
  $q_\mathrm{min} = q_c^\mathrm{Ni}$ in this case.

  This polarizer was characterized using the Selene guide to focus the divergent beam behind the source at the pole position, a $1\un{mm}$ wide slit behind the neutron guide, to an analyzer. 
  The latter was a Si wafer coated with a polarizing $m=6$ SM used in reflecting geometry. 
  The magnetization field was $2000\un{Oe}$ to ensure a high efficiency.
  The spin state of the neutron beam could be switched using an RF flipper located between polarizer and analyzer. 
  In the following we make the conservative (for the calculated performance of the polarizer) assumption that the flipper is 100\% efficient.

\begin{figure}[t]
    \centering
    \includegraphics[width=7cm]{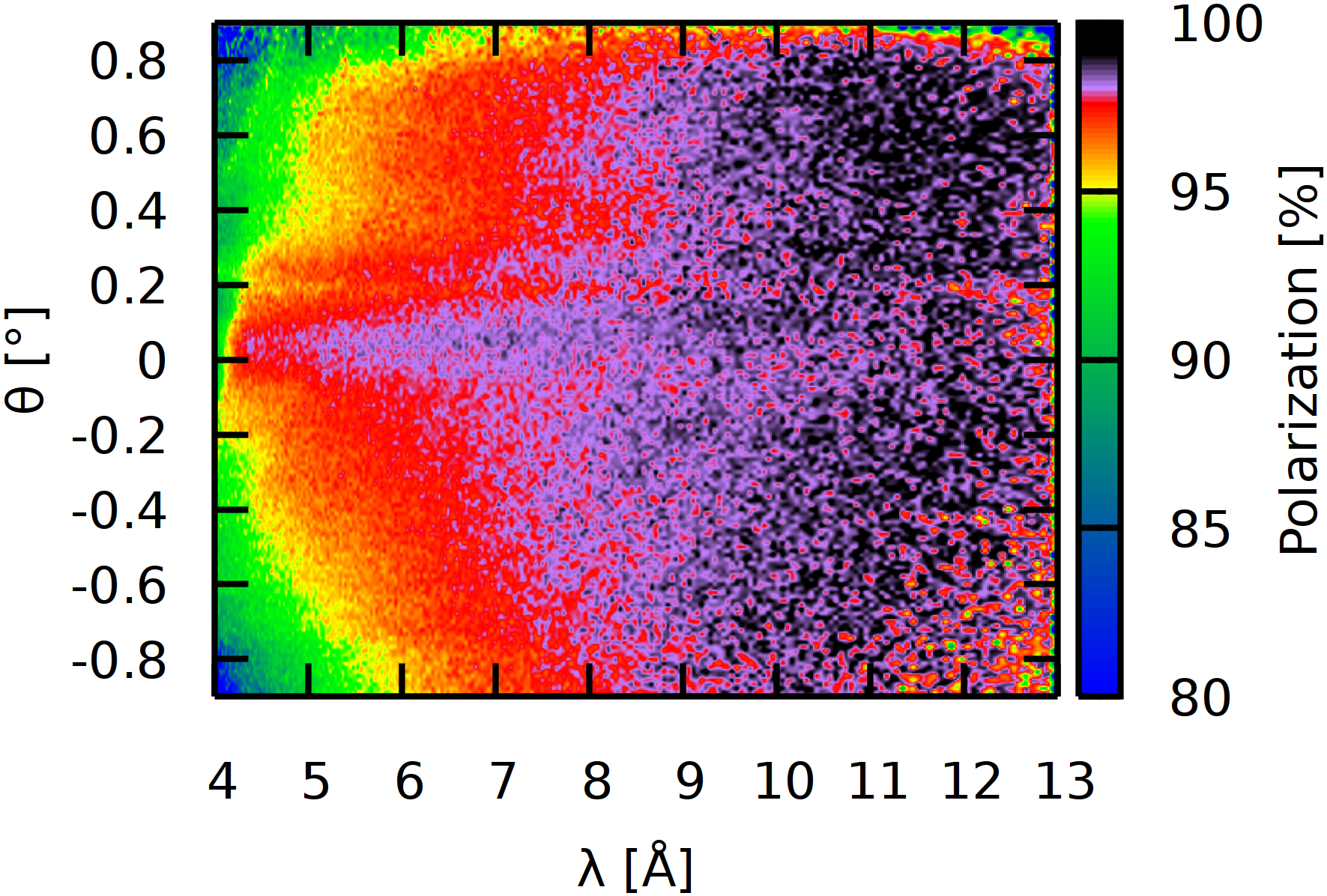} \hspace*{0.5mm}
    \caption{\label{fig:optics_xlnxefficiency}
     Measured polarization efficiency of the demonstrator device, analyzed by a $m=6$ FeCoV/Ti:N supermirror on Si.
     The polarization was flipped with a RF spin-flipper.
     The band in the middle with improved polarization corresponds to transmission through both reflectors.
     Sharp lines visible in the top-left corner originate from the analyzer, while the small reduction around $\theta \approx 0.15^\circ$ is an artifact from the Selene guide transmission.
    }
  \end{figure}

  The efficiency of the polarizer $P_p$ is a function of $\theta$ and $\lambda$, while the efficiency of the analyzer $P_a$ depends on the angle of incidence on its surface, $\omega$, and on $\lambda$. 
  The relative alignment of the analyzer $\omega_a$ determines the relation $\omega = \theta+\omega_a$.
  This can be used to estimate a lower boundary for $P_p$ by performing several measurements with varying $\omega_a$. 
  The measured spin asymmetry $P_{\omega_a}(\lambda,\theta)$ for each pixel is a product of $P_p(\lambda,\theta)$ and $P_a(\lambda,\omega_a+\theta)$.
  Using $P_a\le 1$ this gives $P_p(\lambda,\theta) \ge P_{\omega_a}(\lambda,\theta) \,\forall \omega_a$.
  Figure \ref{fig:optics_xlnxefficiency} shows an intensity map for the lower boundary for $P_p$ obtained from measurements with 4 different $\omega_a$ positions. 
  The gradual decay of $P_a$ with $q$ and its rapid drop at $q_c^\up$ leads to the visible intensity steps in the upper left corner of the map.

  The overall polarization is surprisingly good for only two polarizing transmission interfaces.
  Therefore the device is now used regularly for user experiments at AMOR.
  The darker line for $\theta \approx 0.22^\circ$ corresponds to trajectories traveling through an area of strongly reduced transmission of the Selene guide, which leads to worse signal to noise ratio for the spin-flip channel and thus reduction of the measured polarization.  
  The brighter region at $\theta \approx 0^\circ$ to $0.3^\circ$ corresponds to trajectories intersecting both substrates, i.e. 3 or 4 transmissions through SM interfaces. 
  The lower performance for $\lambda \approx 4\un{\A}$ and $\theta > 0.6^\circ$ or $\theta < -0.6^\circ$, respectively, means that the angle of incidence at the highest $\theta$ angles is larger than expected, which is the result of a too large distance of the polarizer from the source.

  A better alignment (tilt and distance to the source) is expected to deliver
  \begin{eqnarray*}
   P_p &\ge& 93\% \mbox{ for } \lambda > 4\un{\A} \\
       &\ge& 96\% \mbox{ for } \lambda > 7\un{\A}
  \end{eqnarray*}
  over the full $\Delta \theta = 1.8^\circ$, and
  \begin{eqnarray*}
   P_p &\ge& 97\% \mbox{ for } \lambda > 4\un{\A}
  \end{eqnarray*}
  over the range at the device front, where the two mirrors overlap, $\theta = 0.15^\circ$.
  The latter value can also be seen as a lower boundary for the performance of a device with two lamella in sequence.

  The limitations of a single-lamella transmission polarizer are
  \begin{itemize}
    \item $\Delta y/x_{min}$ must be small, typically below $\arcsin 0.35^\circ = 12$
    \item The coating is restricted to $m < 7$, where $P_T$ drops considerably towards this limit, while the costs grow.
          This restricts $\hat{\alpha}$
    \item The wavelength band to be polarized is restricted due to the too high transmission for $\dn$ neutrons at large $q_z$ and total reflection from the substrate for low $q_z$.
    \item For small $\hat{\alpha}$ and large $\lambda$ a parallel offset of the beam is caused by refraction. 
          The offset is in the order of a few ten microns and thus only relevant for very good focusing systems.
  \end{itemize}
 
\section{Multi lamellae devices planned for Estia}
  Estia will be a focusing reflectometer at the European Spallation Source (ESS) that is being build in Lund, Sweden.
  The instrument concept is based on the Selene neutron guide, but in contrast to the focusing option available at AMOR, the elliptical mirrors will form the complete beamline neutron transport system.
  Increasing the size of the focusing neutron optics lead to an increased space available for the polarizing optics.
  
  The goal is to achieve state of the art neutron polarization analysis while keeping the focusing capability and large intensity towards which the instrument is optimized.
  Therefore, larger devices with more than one lamella will be constructed for the beamline. 
  Two separate mirrors for the polarizer will be placed close to the focus in the middle of the Selene guide and a double lamella device on the detector arm as an analyzer.

 \subsection{polarizer}
    \begin{figure}[t]
      \centering
      \includegraphics[width=7cm]{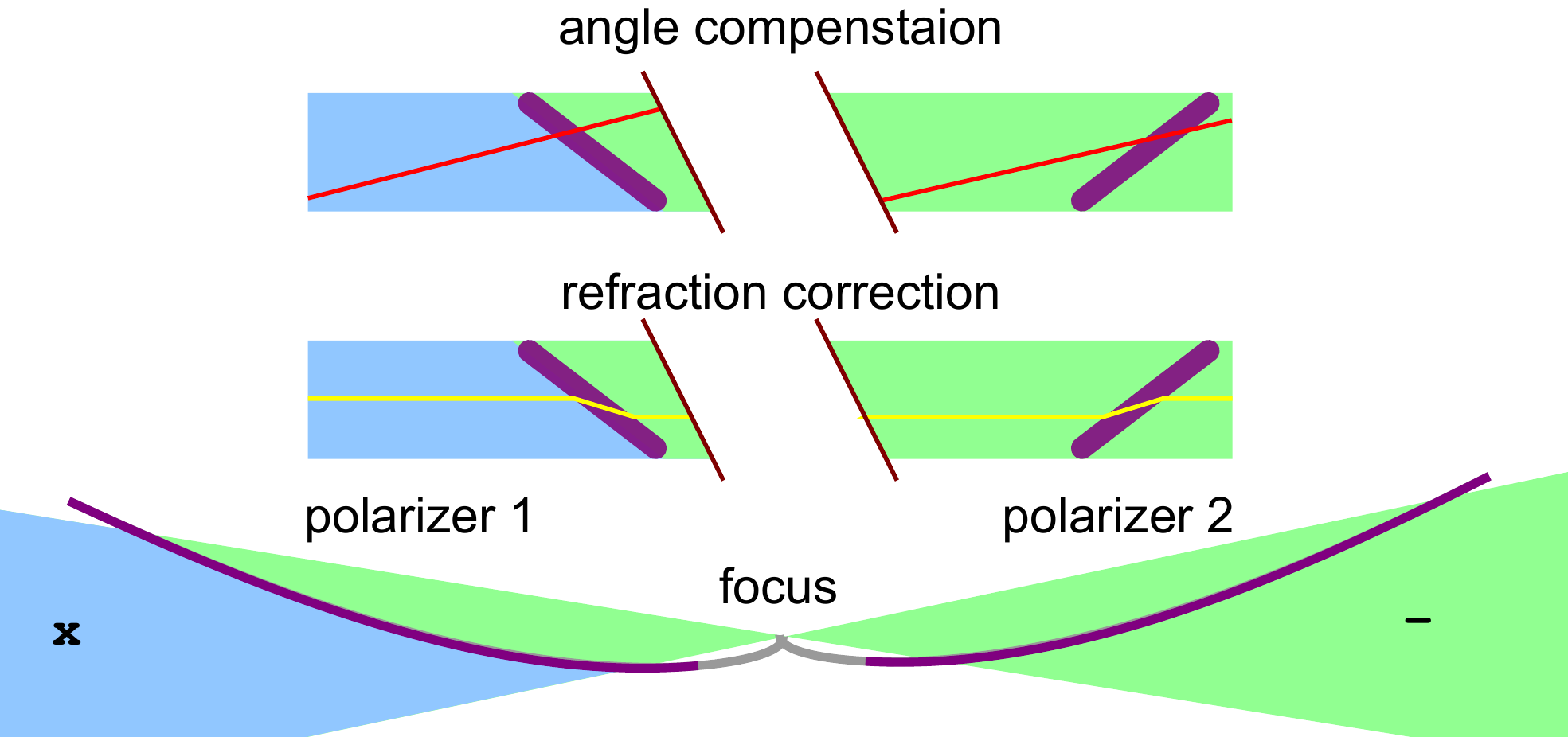}
      \caption{\label{fig:estia_polarizer}
        Polarizer concept for the Estia reflectometer.
        Two double sided supermirrors will be place before and after the focus between the two Selene guides.
        The symmetry of the mirrors has the property that the reflection direction on the second mirror is opposite to that of the first intersection.
        This results in a compensation of polarization efficiency for trajectories that don't go through the focal point as well as a correction of the wavelength dependent beam offset produced by the refraction within the silicon wafer.
      }
    \end{figure}
    Figure \ref{fig:estia_polarizer} shows a conceptual sketch of the Estia polarization concept.
    Two equiangular spirals intersect the beam before and after the focal point in the middle between the ellipses of the Selene neutron guide, where a large free space is available to place these components.
    A large polarization ($>$99\%) and decent transmission\footnote{The transmission of an unpolarized beam is $>$40\% as $>$80\% of spin-down neutrons are transmitted.} ($>$40\%) can be expected from this geometry, as 4 SM interfaces (with m=5.0 Fe/Si) intersect with the neutron path.
    
    The change in intersection direction inherent to the symmetry of the mirror has two additional advantages:
    \begin{itemize}
     \item Trajectories that do not pass through the focal point due to the finite size of the source will hit the second mirror with the inverse angular offset $\delta \alpha$ present at the fist mirror.
          Therefore a too large angle of incident that reduces polarization efficiency of one mirror leads to improved efficiency in the second mirror and vice versa.
     \item The small linear offset due to refraction in the first silicon wafer will mostly be reversed in the second mirror as the angle of incidence has the same magnitude but opposite sign.
    \end{itemize}
   
    With $\hat\alpha=1.65^\circ$ and m=5 the polarization system can cover the whole accessible wavelength range from 4$\un{\A}$ to 25$\un{\A}$ and, at the same time, allow frame overlap suppression above 30$\un{\A}$ by using the silicon total reflection.
    Coatings will be made of Fe/Si in the case of Estia, as the large neutron flux would lead to high activation of the component when using Co containing supermirrors.
    The two devices will be independent, allowing the user to choose between the use of one or two polarizing devices if higher intensity is desired.
    
    \begin{figure}[t]
      \centering
      \includegraphics[width=7cm]{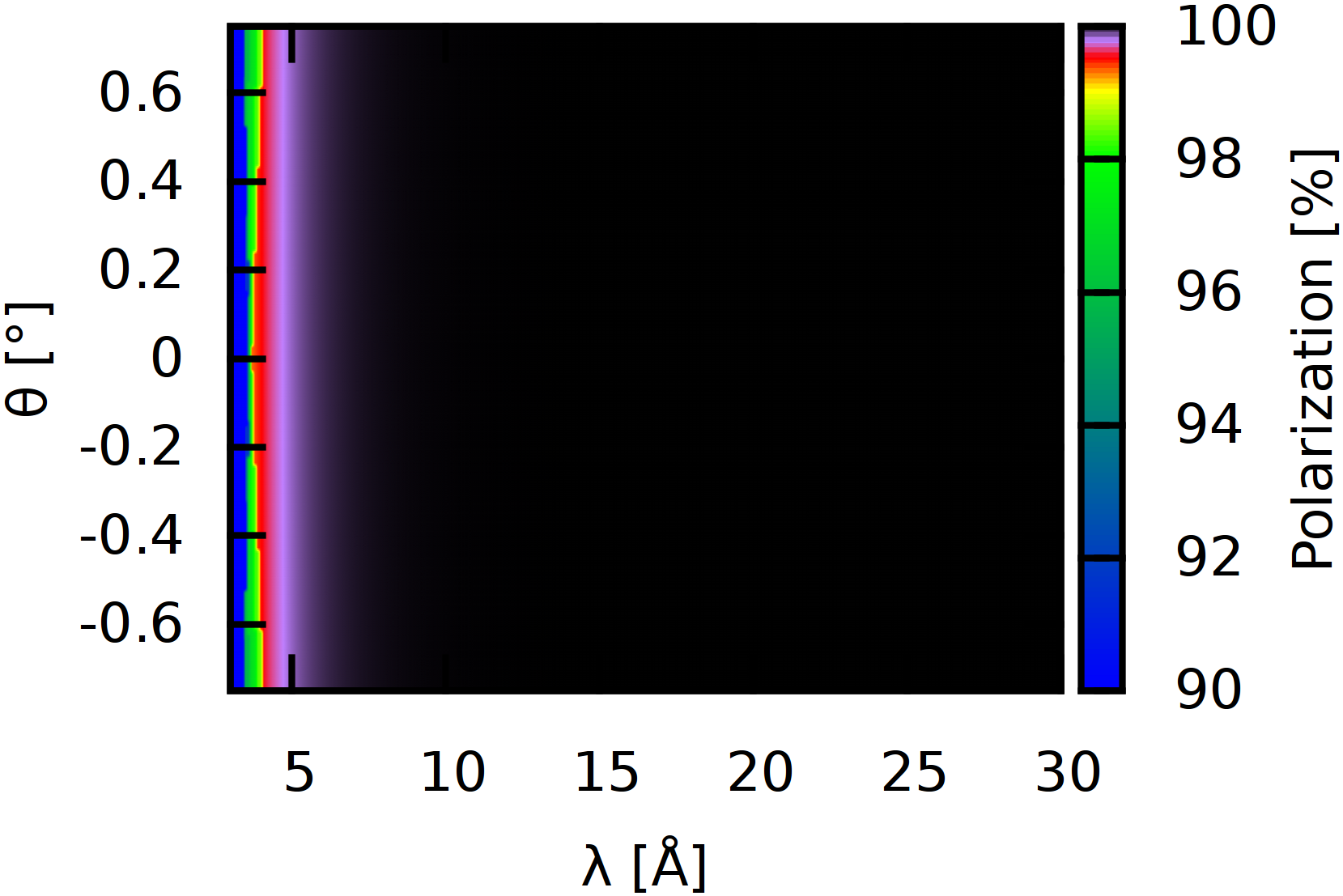}\hspace*{1mm}
      \includegraphics[width=7cm]{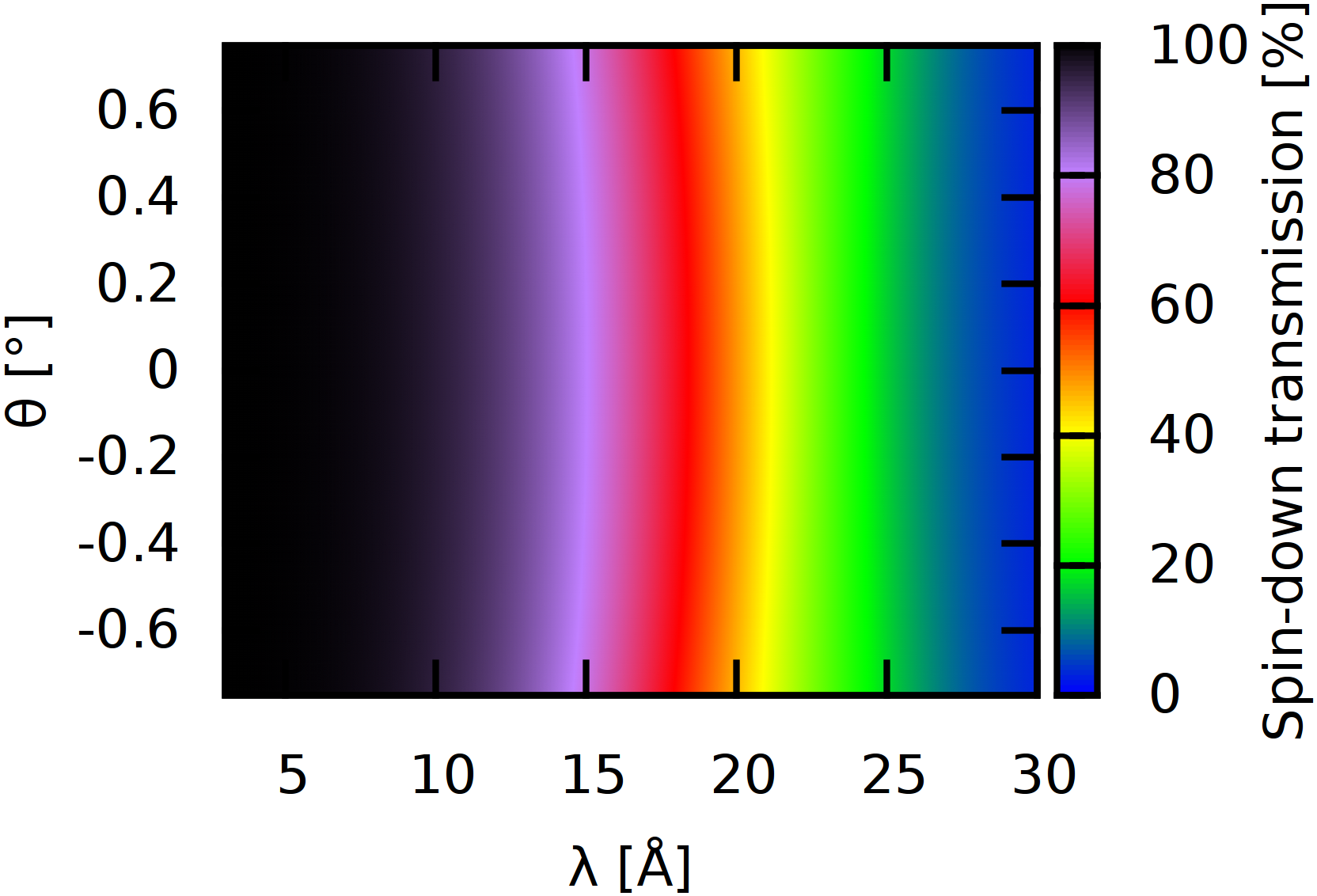}
      \caption{\label{fig:estia_efficiency}
        Calculated polarization efficiency (left) and transmission (right) of Estia double polarizer.
        The primary instrument wavelength band is between 4$\un{\A}$ and 11$\un{\A}$
      }
    \end{figure}

 \subsection{analyzer}
    \begin{figure}[t]
      \centering
      \includegraphics[width=7cm]{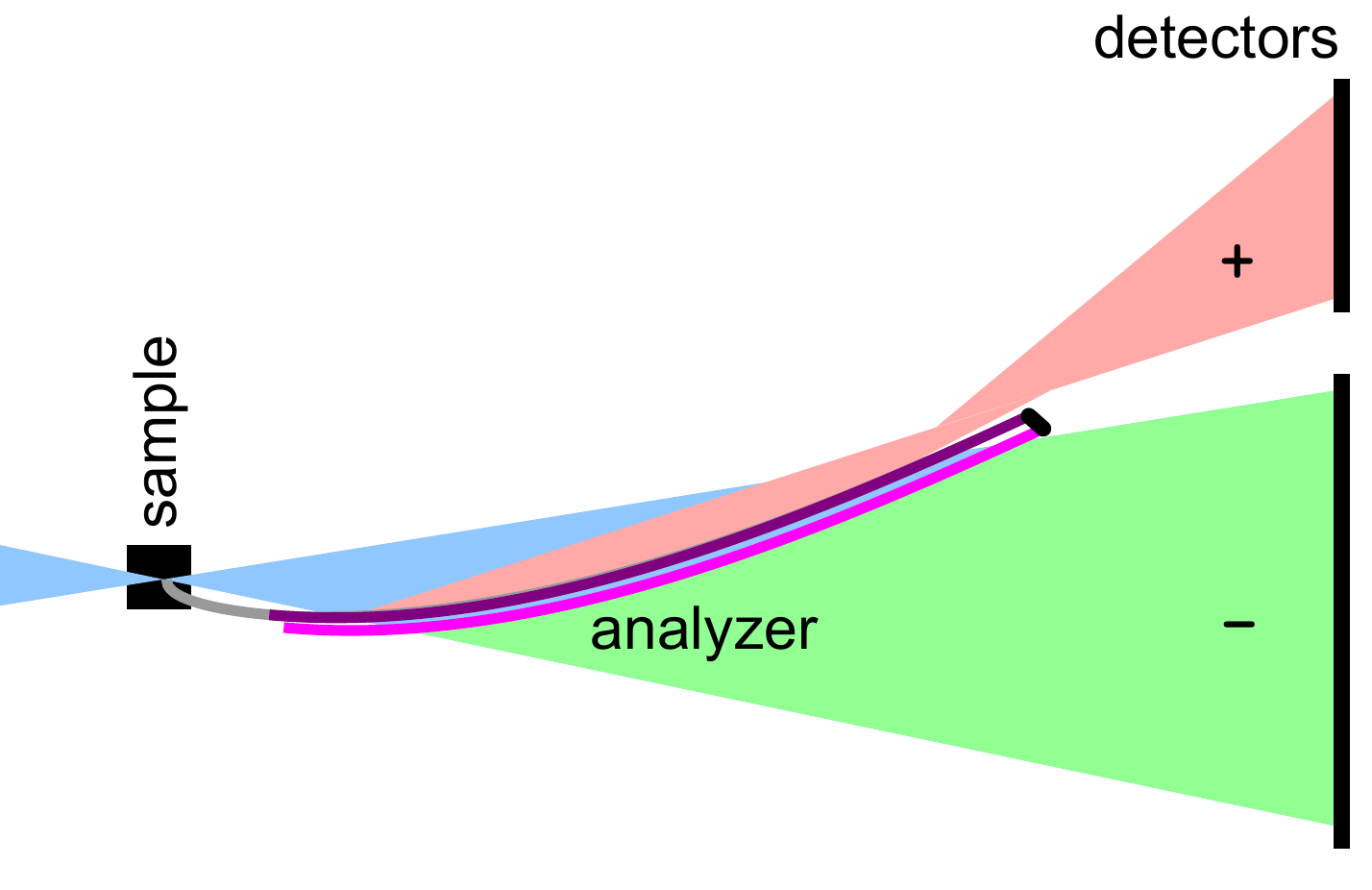}
      \caption{\label{fig:estia_analyzer}
      Analyzer concept for the Estia reflectometer.
      The first mirror is single sided coated as a reflection polarizer while the second is double sided and its reflection is absorbed.
      Capturing both, transmitted and reflected beam allows simultaneous measurement of each spin channel with sufficient polarization efficiency.
      }
    \end{figure}
    Analysis on Estia can not follow the same approach as the polarization, as there is not focal point behind the sample.
    To still be able to reach good polarization efficiency two subsequent transmissions need to be performed.
    When the analyzer spirals are curved in the direction perpendicular to the reflection plane and the geometry (scaling factor $a$) is correctly chosen, it is possible to not only measure the transmitted beam with the position sensitive detector but also the neutrons reflected from the first analyzer mirror.
    
    For this reason the Estia analyzer, which will be placed together with the detector on a movable table, will consist out of two supermirror lamellae (fig. \ref{fig:estia_analyzer}).
    The first will be a m=4 reflection polarizer with just one SM coated side to produce a well polarized spin-up state.
    After this a second, double sided SM with slightly larger $\hat\alpha$ and m-value will be placed, together with an absorber at the end of the device to capture neutrons only reflected on the second lamella.
    As a result the transmitted neutrons will have passed three reflective interfaces before reaching the detector, leading to a relatively good polarization. 
    
    A great gain of detecting both spin-states at the same time can be achieved when analyzing reflectivity from an unpolarized neutron beam, allowing polarized measurements with only slightly reduced intensity (85\%) compared to unpolarized measurements.
    The advantage for polarization analysis is often limited, as the spin-flip signal is much weaker than the non spin-flip reflectivity and dominates the measurement time.
    On the other hand, if two polarized beams, spin-up and spin-down, hit the sample surface at the same time, this method can be used for complete polarization analysis in time resolved measurements.
    A later upgrade to Estia, installing two separate Selene neutron guides which share a common focus at the sample position, will make this simultaneous 4-state measurement possible.

\section{conclusion}
  We have presented a novel concept to use curved polarizing supermirrors in transmission geometry to cover large beam divergences for instruments with beam focusing.
  A demonstration device for the focusing option of the AMOR reflectometer was build and tested on the beamline, showing the expected polarization efficiencies.
  Following the same general concept, the Estia reflectometer at ESS will be build using two spiral shaped polarizers and a double lamella analyzer.
  With this setup the instrument will achieve a high polarization of 99\% over a broad wavelength band with only minor impact on instrument intensity and allow simultaneous measurement of spin-up and spin-down reflectivity on the analyzer side.
  After full extension of the instrument with two separate Selene beam paths, the ability to measure all 4 spin-states at once will allow novel time resolved polarization analysis studies.
  
  As all other polarization devices, the equiangular spiral has its limitations imposed by the SM reflectivity that restricts the range of neutron wavelength and maximum focus size to which the system can be applied.
  With the ongoing advances in pulsed neutron sources and neutron optics, more instruments will have improved focusing capabilities to be able to measure tiny samples.
  In this scenario, current $^3$He systems are unable to deliver sufficient polarization efficiencies or suffer from significant attenuation when applied to beams with typical wavelength bands used at pulsed sources.
  Other solid states devices as e.g. cavities, mirror fans or bender systems significantly distort the neutron trajectories and thus the focusing ability of the instrument (or detector resolution).
  The presented devices are optimized for the focusing geometry and therefore offer significant advantages over the established polarization components.

\section*{References}

\providecommand{\newblock}{}

\end{document}